\documentclass[showpacs,aps,pre,superscriptaddress,twocolumn]{revtex4}

\usepackage{graphicx}
\usepackage{amsfonts}
\usepackage{color, verbatim}

\begin{document}
\title{Bright Solitary Waves on a Torus: Existence, Stability and Dynamics
for the Nonlinear Schr{\"o}dinger Model}
\author{J. D'Ambroise}
\affiliation{ Department of Mathematics, Computer \& Information Science, State University of New York (SUNY) College at Old Westbury, Westbury, NY, 11568, USA}
\author{P.G. Kevrekidis}
\affiliation{Department of Mathematics and Statistics, University of Massachusetts,
Amherst, MA, 01003, USA}
\author{P. Schmelcher}
\affiliation{Center for Optical Quantum Technologies, Department of Physics, University of Hamburg, 
Luruper Chaussee 149, 22761 Hamburg Germany}\affiliation{The Hamburg Centre for Ultrafast Imaging,
University of Hamburg, Luruper Chaussee 149, 22761 Hamburg,
Germany}

\begin{abstract}
  Motivated by recent developments in the
  realm of matter waves, we explore the potential of creating
  solitary waves on the surface of a torus. This is an intriguing
  perspective due to the role of curvature in the shape and
  dynamics of the coherent structures. We find different families
  of bright solitary waves for attractive nonlinearities including
  ones localized in both angular directions,
  %ones localized in the
  %one and periodic in the other direction,
  as well as waves
  localized in one direction and homogeneous in the other.
  The waves localized in both angular directions have also been
  partitioned into two
  types: those whose magnitude decays to zero and those who do not.
   The stability properties of the waves are examined and
  one family is found to be spectrally stable while most are
  spectrally unstable, a feature that we comment on. Finally,
  the nature of the ensuing nonlinear dynamics is touched upon.
\end{abstract}

\maketitle

\section{Introduction}
\label{intro}

The atomic physics platform of Bose-Einstein condensates (BECs)
has served over the past two decades as an excellent testbed
for ideas stemming from nonlinear waves and their interplay
with geometric and topological notions among many
others~\cite{stringari,pethick,siambook}. In particular, a
wide variety of experiments has been performed to explore ideas
related to bright coherent structures in effectively
self-focusing and attractively interacting BECs~\cite{experiment2,expb1,expb2,expb3}
and related to repulsively interacting BECs with their effective self-defocusing nonlinearity
\cite{experiment,experiment1,becker,markus,engels,becker2,markus2,jeff,djf}.
Important structures that were investigated include the gap solitons~\cite{gap}
in optical lattices, multi-component solitons~\cite{revip},
vortices and
multi-vortex configurations~\cite{fetter1,fetter2},
as well as solitonic vortices and
vortex rings~\cite{komineas_rev}.

Typically, these structures are placed in simple confining
potentials with the most canonical example being that of
the parabolic trap~\cite{stringari,pethick,siambook}.
However, this doesn't preclude the possibility that not
only periodic~\cite{markus3}, but also more elaborate non-harmonic
potentials are accessible experimentally, as e.g. described in~\cite{ott}.
In fact, certain experimental techniques enable the
formation of arbitrary trapping geometries~\cite{boshier}
rendering atomic BECs a particularly well suited platform
for exploring the interplay of nonlinearity and geometry.
Nevertheless, other areas too, including nonlinear optics,
provide case examples of different types of confinement
through the manipulation of the optical medium's refractive
index~\cite{kivshar_agr}.

A particular motivation of the present work consists of
a recent theoretical proposal towards exploring optical
lattice settings with a torus topology in the context of BECs,
with the atoms being confined to the surface of a torus~\cite{porto}.
This represents a striking example of a geometrically and
topologically ``nontrivial'' trap which bears great promise
for novel nonlinear wave effects. Indeed there is a plethora
of curved and confining surfaces of different geometry and
topology that might be realizable and accessible experimentally
using modern light manipulation techniques and tools.
In parallel to that, there is an active interest in the field 
of pattern formation towards studying nonlinear
partial differential equations on curved soft substrates;
a prime example of that is the study of defect formation
and localization through a generalized Swift-Hohenberg
theory in elastic surface crystals \cite{reis,dunkel}.
Motivated by an interweaving of these themes, our aim in the
present work is to explore the existence, stability and
dynamics of solitary waves in a nonlinear Schr{\"o}dinger (NLS) model
with attractive inter-particle interactions on the surface
of a torus. This is a prototypical {\it continuum Hamiltonian} case example as opposed to
the lattice one of~\cite{porto} and the dissipative dynamics
of~\cite{reis,dunkel} for the exploration of such dynamics.

Some of our main findings involve the identification of localized
solutions in both angular directions of the torus. Such solutions come in
two families, one on the inside and one on the outside of the
torus. Also solutions depending on only one of the angular variables
(independent of the other) are equally considered and their
families have been identified. Among the doubly localized solutions, we 
find two varieties, one that is more solitonic in nature and one that is more reminiscent of
a periodic solution that does not decay to the proximity
of zero within its spatial variation. This last family, when centered
on the outside of the torus is the only that is found to be spectrally
stable sufficiently close to the small amplitude, i.e. near linear limit.
The dynamics of localized solutions often lead to collapsing spots
on the torus, while those depending on a single angle may lead
to multiple such collapsing spots.

Our presentation is structured as follows. In section II we 
provide the model, the benchmarks used for the linear problem
and the theoretical setup of the nonlinear problem. Then,
in section III, we will examine the different types of nonlinear solutions,
localized and homogeneous ones, that
we have identified in this geometric setting involving the Laplacian
operator on the 2d torus. The continuation of the different branches
of solutions versus the eigenvalue parameter $\mu$ which is the frequency
of the solution, is given and the stability of these states
is considered. When the solutions are identified as unstable
their dynamics will be followed.
Importantly, however, we also identify stable solutions sufficiently
close to the linear limit of the problem.
Lastly, we will summarize
our findings in section IV and present a few of the many directions that are opening
up for this promising research theme.

\section{Model and Theoretical Setup}
\label{thy}

\begin{figure}
\includegraphics[width=1.025\columnwidth]{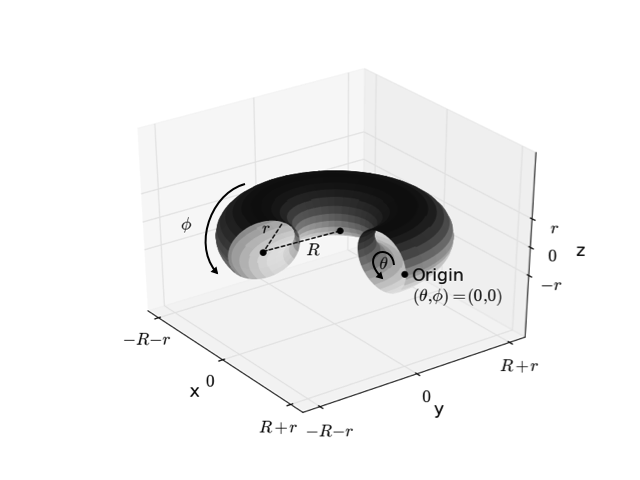} 
\caption{Sketch of a torus with outer radius $R$ and inner radius $r<R$.  The arrows point in the direction of increasing toroidal angle $\phi$ and increasing poloidal angle $\theta$.  }
\label{tordemo}
\end{figure}

{The torus with major radius $R$ and minor radius $r$ is centered at the origin and has parameterization in $\mathbb{R}^3$ given by $x = (R+r\cos(\theta))\cos(\phi)$, $y = (R+r\cos(\theta))\sin(\phi)$, and $z=r\sin(\theta)$.  The toroidal angle is denoted by $\phi$ and poloidal angle $\theta$.  For a diagram of the torus and relevant parameters see Figure \ref{tordemo}.}   Consider the 2D NLS equation on the surface of the torus with radii $R>r$ as follows
\begin{eqnarray}
\label{nls}
i\psi_t  = -\frac{1}{2}\Delta\psi + V(\theta, \phi) \psi - \sigma |\psi|^2\psi 
\end{eqnarray}
where $\sigma = 1$ is the focusing and $\sigma=-1$ the defocusing case.  We will focus on the former setting in the present study.  The Laplacian operator takes the form:
\begin{eqnarray}
  \Delta =  \frac{1}{r^2}\partial_{\theta\theta} - \frac{\sin\theta}{r(R+r\cos\theta)} \partial_\theta + \frac{1}{(R+r\cos\theta)^2}\partial_{\phi\phi}.
  \label{lap}
  \end{eqnarray}
In what follows, we
define $\alpha = r/R$; moreover, as a starting point we will
hereafter set $V(\theta, \phi)=0$ to examine the potential of
the torus for intrinsically localized states.
A natural benchmark that we have performed
en route to the consideration of the nonlinear problem
has been the study of the linear spectrum of the underlying Laplacian
operator as identified, e.g., in~\cite{glowinski}. This
spectral analysis has been done as a function
of the parameter $\alpha$ characterizing the nature of the torus
as regards the size of the minor radius $r$ over the major radius $R$.
A key feature in this case is that the relevant  spectrum is discrete and consists
of isolated eigenvalues due to the presence of a finite interval for the
values of $\theta$ and $\phi$ (both running in the $[0,2 \pi]$
interval).

As is customary in the NLS setting, we consider
stationary solutions by setting $\psi = e^{-i\mu t}u$ and obtaining the
boundary value problem naturally with periodic boundary conditions
\begin{eqnarray}
\label{stateq}
-\frac{1}{2}\Delta u + V(\theta, \phi) u - \sigma |u|^2u -\mu u = 0.
\end{eqnarray}
Once the solutions (among them we are interested in the ones
that bear some form of localization)
to Eq.~(\ref{stateq}) have been identified,
their spectral stability can be monitored by setting
\begin{eqnarray}
  \psi &=& (u + \delta\left[a(\theta,\phi)e^{\nu t} + b(\theta,\phi)^* e^{\nu^* t}\right])e^{-i\mu t}.
  \label{perturb}
\end{eqnarray}
From this,  we obtain to order $\delta$ the linear system
\begin{equation}
\left[ \begin{array}{cc} M_1 & M_2 \\ -M_2^* & -M_1^* \end{array} \right] \left[ \begin{array}{c} a\\ b \end{array} \right] =  -i\nu \left[ \begin{array}{c} a\\ b \end{array} \right]\label{mat}
\end{equation}
where $M_1 = \frac{1}{2} \Delta - V  + \mu + 2\sigma |u|^2$ and $M_2 = \sigma u^2$.   From Eq.~(\ref{perturb}) it follows that
max(Re$(\nu)$) $> 0$ corresponds to instability.

{
  We focus primarily on the stationary solutions with some kind of localization.  Such solutions are found to have the largest bulk of the
  {waveform} located either on the inside or the outside of the torus.  These special locations on the torus correspond to where $\sin(\theta)=0$.
  %and thus the middle term of the Laplace operator in (\ref{lap}) is extremized.
  Next we
  separate categories of solutions based on whether the bulk of the solutions is centered near $\theta\approx 0$ on the outside of the torus (Types -out) or $\theta\approx \pi$ on the inside of the torus (Types -in).  Note also that the operator in (\ref{lap}) remains translationally invariant in $\phi$ and so correspondingly we find some solutions which  appear as a solid stripe wrapping around the torus in the toroidal $\phi$ direction.
}

%The stationary equation (\ref{stateq}) can be solved via Newton's method by using various initial guesses.  We will use the following classification to describe solutions according to a similar-looking initial guess from which the solutions can originate. 
{The classification of solutions below is also based on which directions the solutions are localized.  Using Newton's method to solve the stationary equation (\ref{stateq}),  one can select an initial guess
  bearing the desired localization properties of an intended
  solution in order to obtain convergence of the iterative process.} {We focus on solutions which have a single concentration of mass.}  All of the solutions which we have identified for the nonlinear problem with $\sigma=1$ in the context of the present study
lie on a spectrum between the category types outlined below,
and have been obtained via continuation
in the frequency parameter $\mu$.

A Type I solution {has localization properties} similar to an initial guess in the form of 
\begin{eqnarray}
u= A_0 {\rm sech}\left( \sqrt{|\mu| \left[ B_0(\theta-\theta_0)^2 + C_0(\phi-\phi_0)^2 \right]} \right) \label{ellipseguess}
\end{eqnarray}
 {for constants $A_0, B_0, C_0\in \mathbb{R}$.  For sufficiently large $|\mu|$} the solutions are localized both in the
 toroidal and in the poloidal direction.  {As $|\mu|$ transitions to smaller values the solution footprint grows and comes to wrap around the torus in both directions.}
 
Such solutions appear as a two-dimensional sech-shape in both the $\theta$ and $\phi$ directions on the surface of the torus.  There are two primary subtypes.  {We will call a Type I solution Type I-in if its bulk is on the interior of the torus; such solutions can be obtained from an initial guess of the form (\ref{ellipseguess}) with $\phi_0,\theta_0\approx \pi$.  Type I-out solutions bear mass predominantly on the exterior of the torus and they can be obtained from an initial guess (\ref{ellipseguess}) with $\phi_0 \approx 0$, $\theta_0 \approx 0$.}
{One of these configurations will correspond to a local
  energy minimum, while the other will correspond to a nonlinear
  Hamiltonian energy maximum
  though which one is which also depends on the value of $\alpha$.}

%We will call a solution {\color{blue} with similar localization features as} (\ref{ellipseguess}) with  $\phi_0,\theta_0\approx \pi$ a Type I-in solution since they are positioned on the inside of the torus surface, and with $\phi_0 \approx 0$, $\theta_0 \approx 0$ a Type I-out solution since such solutions are positioned on the outside of the torus surface.

Type II solutions have localization properties similar to either an initial guess of the form
\begin{eqnarray}
u =  {\rm sech}(A_0(\phi-\phi_0))(1+B_0 \cos(\theta-\theta_0)) \label{cosguess}
\end{eqnarray}
 {for constants $A_0, B_0 \in \mathbb{R}$, or the same format but with $\theta, \phi$ switched. These solutions are localized in the one direction and wrap around
   the torus (but without approaching zero)
   in the other direction.}  There are once again
 two sub-types.  Type II-in {solutions bear most of their mass
   on the interior of the torus and can be obtained from initial guess (\ref{cosguess})} with $\theta_0, \phi_0\approx \pi$, and Type II-out {solutions are principally localized on the exterior of the torus and can be obtained from initial guess (\ref{cosguess}) with} $\theta_0 \approx 0$, $\phi_0 \approx \pi$. 
 Recall, however, that type I and type II solutions can equivalently
 be localized at any other value of $\phi_0$, given the translational
 invariance of the Laplacian operator on $\phi$.
 
Finally, Type III solutions have localization properties similar to an initial guess of the form
\begin{eqnarray}
u =  {\rm sech}(A_0(\theta-\theta_0)) \label{stripeguess}
\end{eqnarray}
{for $A_0\in\mathbb{R}$ a constant}.  {Such  resulting} solutions appear as a localized shape in the poloidal $\theta$ direction and a solid stripe, i.e. it does not depend
on and is thus uniform in the
toroidal $\phi$ direction.
{The Type III-in solutions are approximately centered at $\theta_0\approx\pi$ and the Type III-out at $\theta_0\approx 0$.}

As a diagnostic computed when constructing the bifurcation diagrams
of the different states, we use
the power of stationary solutions in the form {of the following surface integral on the torus}:
\begin{eqnarray}
  P = \int_0^{2\pi}\int_0^{2\pi} |\psi|^2 dS
  \label{norm}
\end{eqnarray}
{where the surface element on the torus is $dS= |\hat{\phi}\times\hat{\theta}| d\theta d\phi$.}  Here hats denote the unit vectors in the different (toroidal and poloidal) directions. Thus, $|\hat{\phi}\times\hat{\theta}| = R + r\cos(\theta)$.
Figures \ref{newt_15_1}-\ref{newt_75_2} show examples of stationary solutions with their magnitude  shown
in color on the surface of the corresponding torus.  The power
(used as a bifurcation parameter for the solution branch)
and the maximum real part of the eigenvalues,
identifying the spectral stability of the solution, are shown as a function of $\mu$.  Figure \ref{flatplots} shows examples of stationary solutions with the magnitude of solutions shown in terms of $\theta$ versus $\phi$ in the two-dimensional flat rectangle $[0,2\pi]\times [0,2\pi]$.  Figures \ref{rk_75mu-8}-\ref{rk_5} show the dynamics of unstable solutions.
  These results will be discussed in more detail in the next section.

{
  It is relevant to note that additional
  wave solutions of Eq. (\ref{stateq}) exist and they can be found using an initial guess of $e^{il \phi}$ for $l=1, 2, \dots$.  Such solutions are unstable with the fundamental (non-vortical) states of $l=0$ being more dynamically
  robust than those with $l \neq 0$ for the cases we have considered.
  All examples observed for $l \neq 0$ feature blowup similar to the fate of the localized solutions described in the next section.  
}

\section{Numerical Results}

\begin{figure}
\includegraphics[width=1.025\columnwidth]{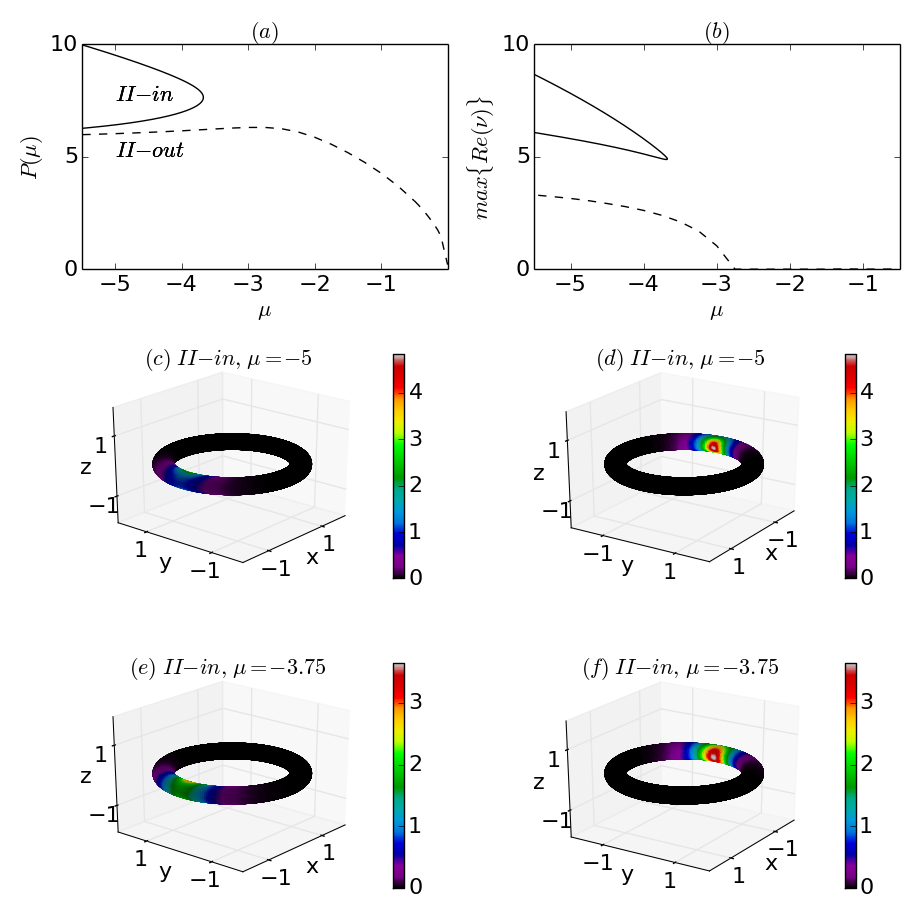} 
\caption{These graphs correspond to $\alpha = 0.15$ and $R = 1.6$ so that $r = \alpha R = 0.24$.  In (a) the power $P(\mu)$ of solutions is shown as a function of $\mu$.  In (b) the maximum real part of the eigenvalues $\nu$ is shown as a function of $\mu$.  The magnitude of stationary solutions $|u|$ is shown according to the colorbar on the surface of the torus.  A solution with $\mu=-5$ in subfigures (c) and (d) (two views, front and back) is of Type II-in, and although the mass of the solution is principally
  located on the interior of the torus a small amplitude does wrap around in the poloidal direction.   A continuation of this solution in the parameter $\mu$ gives the other Type II-in solution with  $\mu = -3.75 $ in subfigures (e) and (f) (two views, front and back) that is localized in the toroidal direction and wraps around in the poloidal direction.  {As the branch continues and the power curve turns upwards to greater power values the solutions continue to broaden in the toroidal direction.} The power and eigenvalues for these solutions and others obtained by continuation in $\mu$ are shown in the top panels with a solid line.  Examples of solutions in Figure \ref{newt_15_2} correspond to the other branch shown with a dashed line.  The latter branch is spectrally stable if
$\mu$ is sufficiently large (and its absolute value sufficiently small).}
\label{newt_15_1}
\end{figure}

\begin{figure}
\includegraphics[width=1.025\columnwidth]{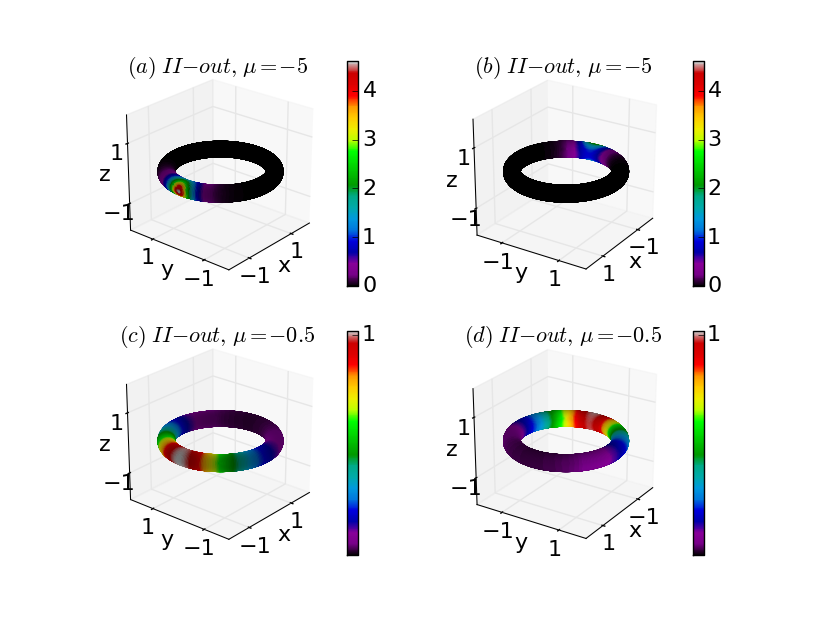} 
\caption{These graphs on the torus are similar to those seen in Figure \ref{newt_15_1}.  The solutions here correspond to the power and eigenvalue curves from Figure \ref{newt_15_1} with a dashed line.  The solution with $\mu=-5$ in subfigures (a) and (b) (front and back views) is of Type II-out, having its bulk on the exterior of the torus but also wrapping around in the poloidal direction.  This solution originates from continuation in $\mu$ from the other solution of Type II-out in subfigures (c) and (d) (front and back views) which has value $\mu = -0.5 $.  The solution with $\mu=-5$ has more mass located on the exterior of the torus.  }
\label{newt_15_2}
\end{figure}

We now turn to the presentation of our numerical findings.
We will consider different values of $\alpha$ progressing from
smaller to larger ones. 
In Figures \ref{newt_15_1} and \ref{newt_15_2} two branches of solutions are shown for $\alpha = 0.15$.
One branch (solid line in Figure \ref{newt_15_1} (a)-(b)) has Type II-in solutions  that are shown in Figure \ref{newt_15_1} (c)-(f).  These solutions are sech-shaped in the toroidal direction and {wrap around} in the poloidal direction (without getting
close to zero), with the bulk of the solution located on the interior of the torus.   For larger values of $|\mu|$  the solutions have a smaller footprint on the torus, i.e. they become more localized as we
expect in this setting of stronger nonlinearity.  For smaller values of $|\mu|$ on this branch the solution footprint is broader on the torus.  This branch has increasing power until $\mu \approx -3.5$ where the branch
{has a turning point and continues to higher power as $\mu$-values decrease
(increase in absolute value)}.
The other branch (dashed line) has Type II-out solutions that are shown on the torus in Figure \ref{newt_15_2} (a)-(d).  These solutions are similar functionally but with the bulk of the solution located on the exterior of the torus.  This branch has increasing and then decreasing power as a function of $\mu$.  {This change in the power's monotonicity is caused by a change in the number of real eigenvalues, i.e., a change {in stability}.
  There are six real eigenvalues for higher values of $|\mu|$.
  Four are zero, and two have symmetric nonzero real values accounting for the instability
  of the branch.   Around $\mu \approx - 2.75$ the count of real eigenvalues changes to four (all are zero) and the solutions spectrally stabilize as the previously mentioned two nonzero real eigenvalues now merge onto the imaginary axis.} 

\begin{figure}
\includegraphics[width=1.025\columnwidth]{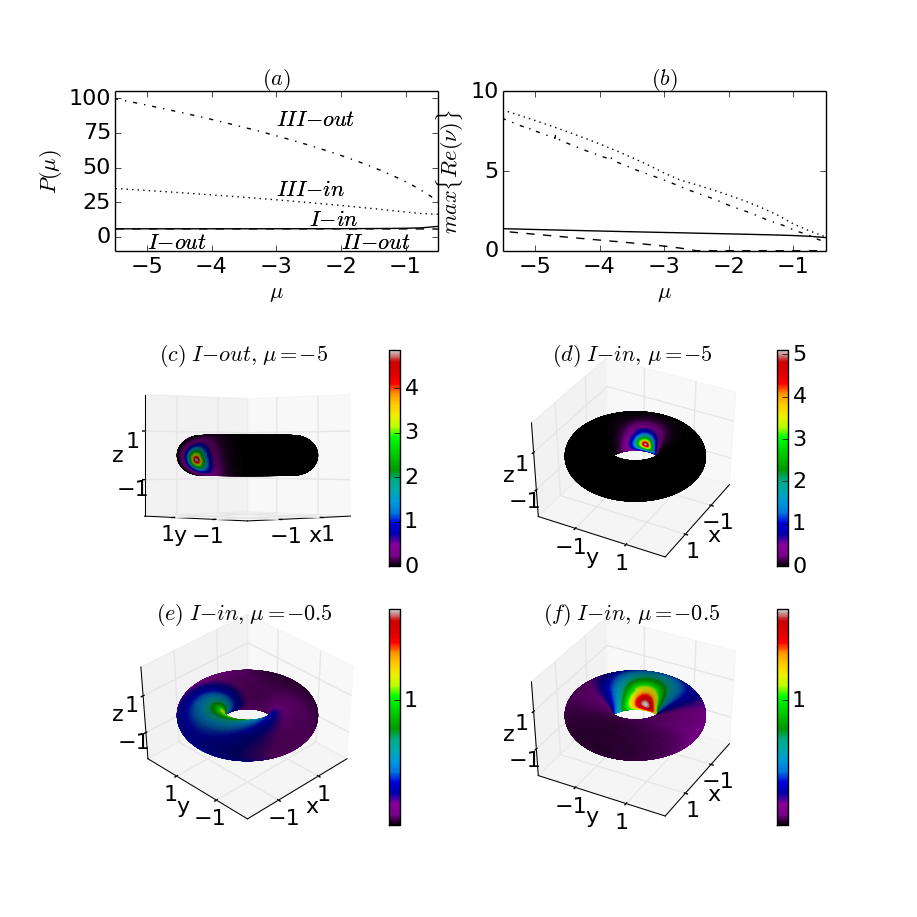} 
\caption{As in Figure \ref{newt_15_1} but for the values $\alpha = 0.5$, $R = 1.6$, and $r = \alpha R = 0.8$.   The solid lines in subfigures (a) and (b) correspond to Type I-in solutions which look as a sech shape in both directions.  Examples on this branch are shown in (e) and (f)  $\mu = -0.5$ (front and back views) and also in (d) for $\mu = -5$.  The dashed lines in subfigures (a) and (b) correspond to Type I-out solutions which are sech shape on the exterior of the torus and after continuation to lower values of $|\mu|$ transition into Type II-out solutions which {wrap around} in the poloidal direction.  In (c) an example is shown on the dashed branch for $\mu = -5$ which is sech shaped in both directions. }
\label{newt_50_1}
\end{figure}

\begin{figure}
\includegraphics[width=1.025\columnwidth]{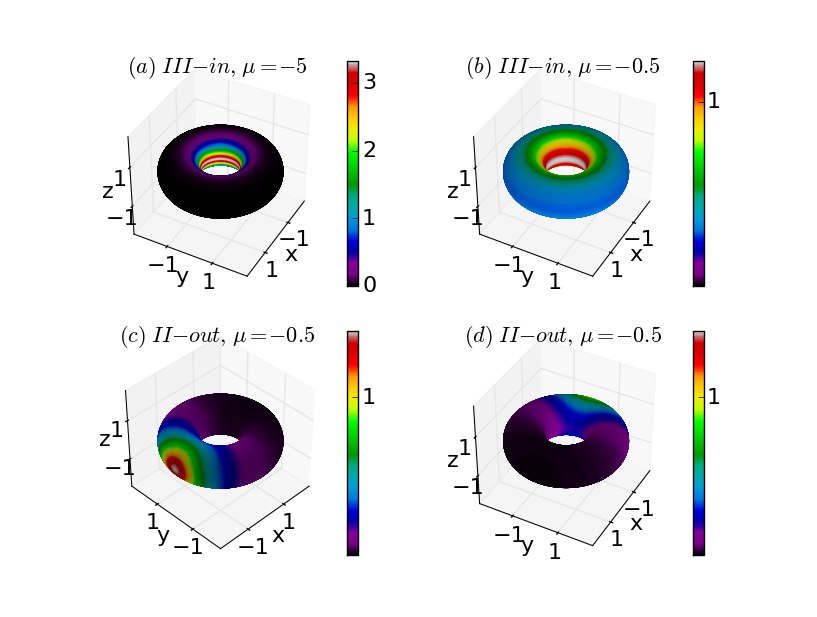} 
\caption{These graphs on the torus are similar to those provided in Figure \ref{newt_50_1}.  The solutions subfigures (a) and (b) are of Type III-in and correspond to the dotted line branch from Figure \ref{newt_50_1}.   For these Type III-in solutions the stripe visible on the interior of the torus is solid around the hole of the torus due to the independence of the solution on $\phi$.  The solution shown here in (c) and (d) (front and back views) is a Type II-out example with $\mu = -0.5$ {wrapping around} in the poloidal direction.  This lies on the dashed branch from Figure \ref{newt_50_1}.   }
\label{newt_50_2}
\end{figure}

In Figures \ref{newt_50_1}-\ref{newt_50_2} {four} branches of solutions are demonstrated with $\alpha = 0.5$.  One branch (solid line in Figure \ref{newt_50_1} (a)-(b)) has Type I-in solutions which are sech shaped in both directions. Here too, lower $|\mu|$ leads to a
wider footprint, while higher $|\mu|$ has smaller footprint on the surface of the torus. The latter also leads to higher intensity in this more highly
nonlinear regime.  The Type I solutions are shown in Figure \ref{newt_50_1} (c)-(f). Another branch (dashed line) has Type I-out solutions which are {localized}  in both directions for higher $|\mu|$ values, and when they are continued in the $\mu$ parameter to lower $|\mu|$ values they widen {and eventually wrap around} in the poloidal direction to approach Type
II-out solutions. {Other branches (dotted and dash-dotted line) are found here for $\alpha = 0.5$ in which solutions are of Type III-in and III-out as a solid stripe around the interior or exterior of the torus respectively.}  Generally these Type III branches are found to exist for other values {such as  $\alpha \gtrsim 0.45$} too. These branches typically carry
larger mass, %(given their homogeneous along $\phi$ nature),
as the solitary structure is quasi-one-dimensional
being uniform along the toroidal direction.   The Types II and III examples are shown in Figure \ref{newt_50_2} (a)-(d).

\begin{figure}
\includegraphics[width=1.025\columnwidth]{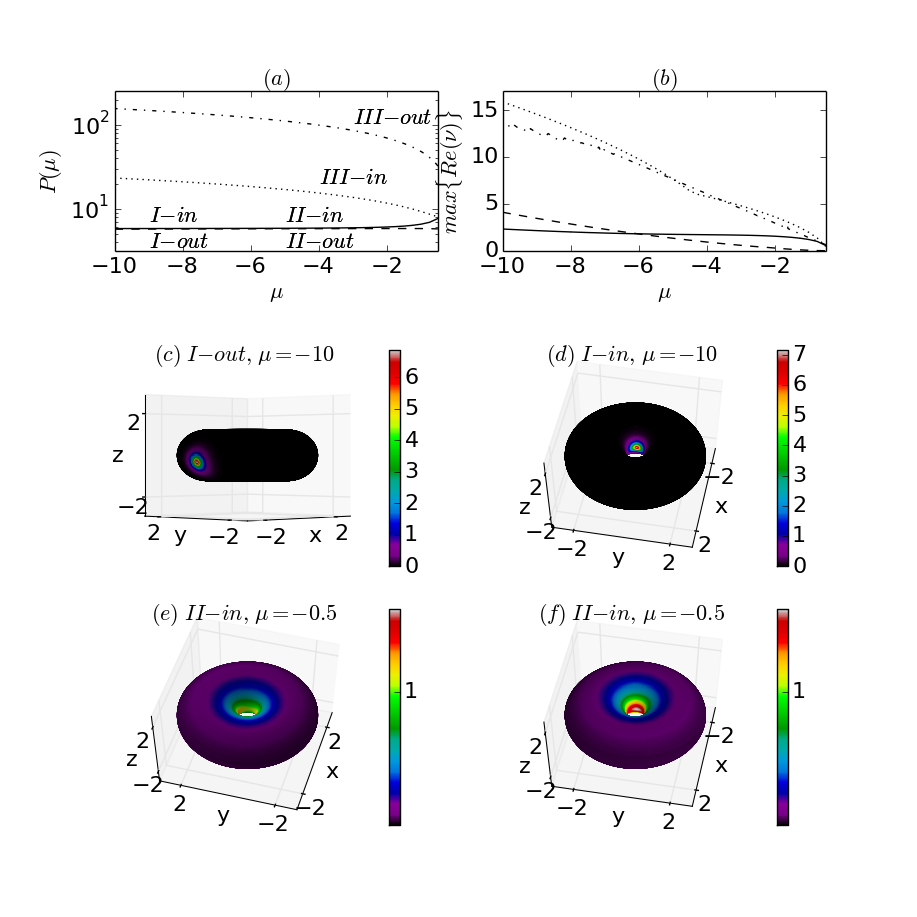} 
\caption{These graphs are similar to Figure \ref{newt_15_1} but here with values $\alpha = 0.75$, $R = 1.6$, and $r = \alpha R = 1.2$.   {Since the power becomes large for some branches in (a), this graph is displayed with a log scale on the y-axis.}  The solid lines in (a) and (b) correspond to localized
  in both directions Type I-in solutions for higher $|\mu|$ values which upon continuation to lower $|\mu|$ values give Type II-in solutions {as they widen in the toroidal direction}.  Examples on this solid line branch are seen in subfigures (e) and (f) for $\mu = -0.5$ (front and back views) and also in (d) for $\mu = -10$.  The dashed lines in (a) and (b) correspond to Type I-out solutions for higher values of $|\mu|$ which transition into Type II-out solutions for lower values of $|\mu|$.  In (c) an example is shown on this dashed branch for $\mu = -10$.}
\label{newt_75_1}
\end{figure}

\begin{figure}
\includegraphics[width=1.025\columnwidth]{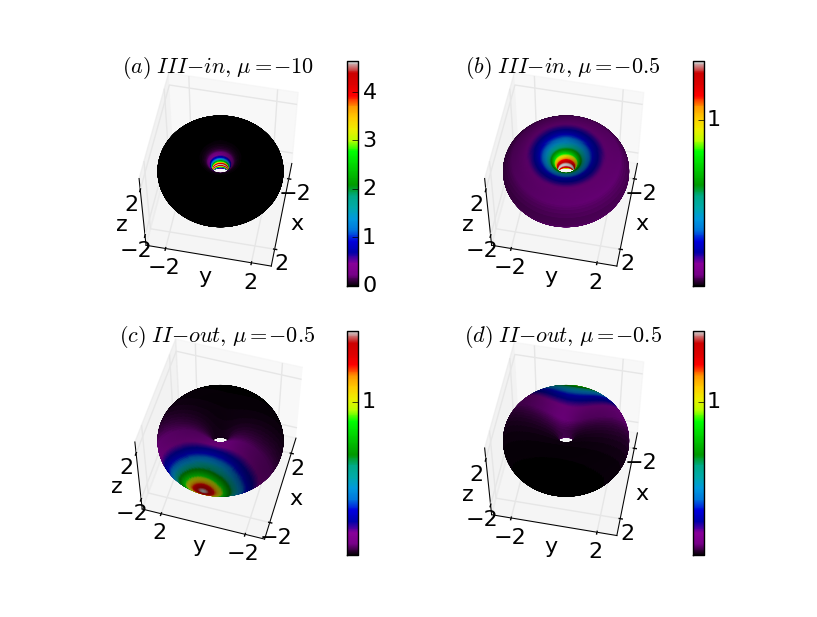} 
\caption{These graphs on the torus are similar to those seen in Figure \ref{newt_75_1}.  The solutions in (a) and (b) are of Type III-in and correspond to the dotted line branch from Figure \ref{newt_75_1}.  For these Type III-in solutions the stripe visible on the interior of the torus is solid around the hole of the torus given the homogeneous nature of the solution in the corresponding ($\phi$)  variable.  The solution here in subfigures (e) and (f) (front and back views) is a Type II-out example with $\mu = -0.5$ which lies on the dashed branch from Figure \ref{newt_75_1} {and it wraps around in the poloidal direction}.   }
\label{newt_75_2}
\end{figure}

In Figures \ref{newt_75_1}-\ref{newt_75_2} again {four} branches are found to exist for $\alpha = 0.75$.  One branch (solid line in Figure \ref{newt_75_1} (a)-(b)) has Type I-in solutions which are sech-shaped in both directions for large $|\mu|$.    As $|\mu|$ decreases, the solution widens and wraps around {in 
  the toroidal direction}.
This effectively implies that Type I solutions morph (for
this value of $\alpha$) into Type II solutions, upon the relevant
variation of $\mu$.  
{This is to be contrasted
  with the  $\alpha=0.5$ (thinner torus) case described above where the solutions come to wrap around in both directions as $|\mu|$ decreases.  For the $\alpha = 0.75$ case (fatter torus) the broadening of the solutions occurs only in the toroidal $\phi$ direction. }  Another branch
  (dashed line) is similar to the dashed line for the $\alpha = 0.5$ case
  described above where Type I-out solutions transition to Type II-out
  solutions {by widening and wrapping around in the poloidal direction}.  See Types I and II solution shown in Figure \ref{newt_75_1} (c)-(f) and in Figure \ref{newt_75_2} (c)-(d).  The
  third {and fourth} branches which are of Type III solutions shown in Figure \ref{newt_75_2} (a)-(b)
  are similar to the $\alpha = 0.5$ case and they exist for 
  other values of {$\alpha \gtrsim 0.45$}. It is relevant to
  note in passing that all of these solutions for the larger values
  of $\alpha$ have been found to be unstable. 

\begin{figure}
\includegraphics[width=1.025\columnwidth]{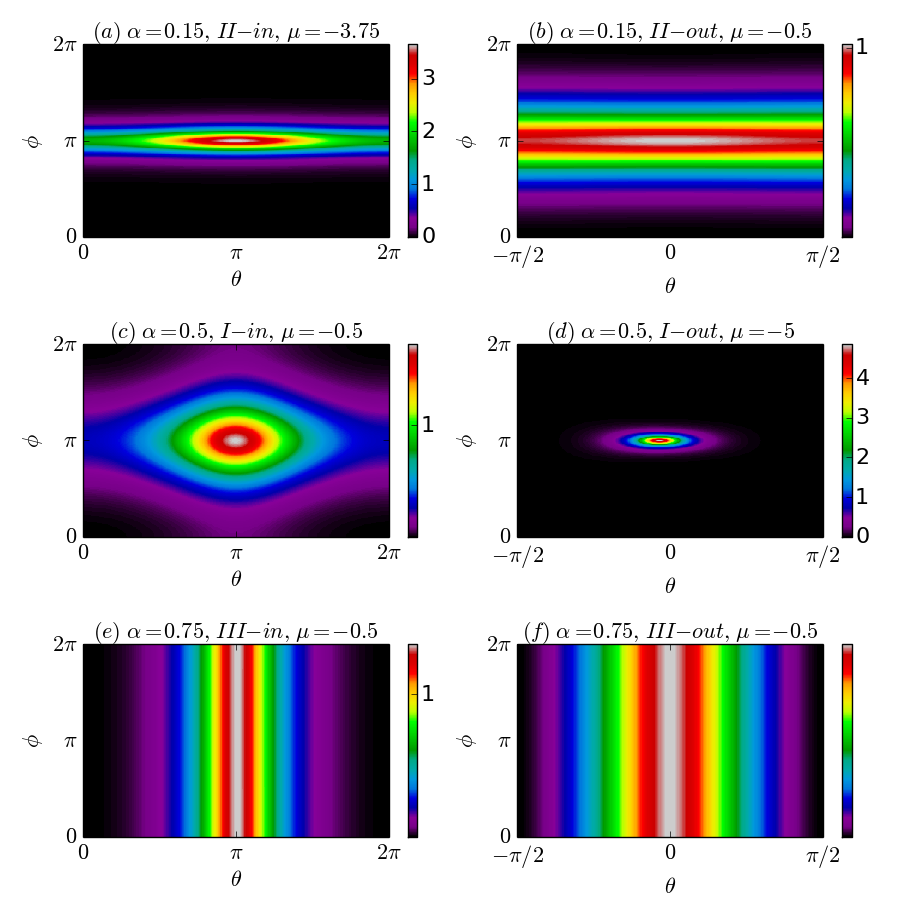} 
\caption{The graphs demonstrate some of the different shapes of solutions, represented flat with respect to $\theta, \phi \in [0, 2\pi]$.  Plot (a) here is the same solution as (f) in Figure \ref{newt_15_1}.  Subfigure (b) here is the same solution as (d) on Figure \ref{newt_15_2}; since this is an out-type solution the flat axis here is centered at the peak of the solution.  Subfigures (c) and (d) here are the same solutions in (f) and (c) in Figure \ref{newt_50_1} respectively.  {Subfigures (e) and (f) here are Type III-in and Type III-out respectively for $\alpha = 0.75$ at $\mu = -0.5$.  }  }
\label{flatplots}
\end{figure}

{
  In Figure \ref{flatplots} the magnitude of some solutions is shown
  in a flat rendering on the rectangle $[0,2\pi]\times [0,2\pi]$ with respect to $\theta$ and $\phi$ to give a ``planar'' sense of the waveforms.  The top row (a)-(b) shows typical Type II solutions which wrap around in the toroidal direction
  (without returning to the proximity of zero)
  and are localized in the poloidal direction.  The middle row (c)-(d) shows a standard
  example of a Type I solutions localized in both directions for sufficiently large values of $|\mu|$. Upon parametric continuation in $\mu$ the solution footprint expands in both directions so that for small enough values of $|\mu|$ solutions wrap around in both toroidal and poloidal directions.  The bottom row (c)-(f) shows
  some basic examples of Type III solutions which represent a solid stripe in the toroidal direction.
}

According to the panels (b) in Figures \ref{newt_15_1}, \ref{newt_50_1},
\ref{newt_75_1}
which summarize
the stability features of the obtained solutions, all the branches
identified as having stable solutions for a range of frequencies
are of Type II-out, i.e., they are localized in the toroidal direction
but do not approach zero along the poloidal direction,
and they have most of their mass on the outside of the torus.
Notice that this happens only for an interval of $\mu$,
indicating that this stability is a byproduct of the interplay
between the nonlinearity and geometry. For the two-dimensional
plane in the absence of curvature,
recall that solitary wave two dimensional solutions
are prone to collapse as is discussed in detail e.g. in~\cite{sulem,fibich}.
On the other hand, the geometry breaks the translational invariance
of the model, as well as the scale invariance thereof, leading
to the existence (for sufficiently strong nonlinearity) of real
eigenvalue pairs. It is only for this (II-out) branch of solutions
and for a range of frequencies in the vicinity of the small
amplitude limit that the interplay of curvature and nonlinearity
permits spectral stability.
Such stable solutions exist for a wider range of $\mu$-values if $\alpha$ is
small (on a thinner torus), and they exist for a smaller range of
$\mu$-values for $\alpha$ large (on a fatter torus).  Plots of stable
solutions are shown in plots (c) and (d) in Figures \ref{newt_15_2}, \ref{newt_50_2}, \ref{newt_75_2} for $\alpha = 0.15, 0.5, 0.75$ respectively.  By comparing these stable solutions one can see that the amount of mass of these stable solutions which wraps around in the
poloidal direction starts to diminish as $\alpha$ increases.

It is relevant to note that from the stability perspective, solutions
of Type I and Type II possess two pairs of eigenvalues at the origin,
reflecting the invariance with respect to phase and the translational
invariance along the toroidal $\phi$ direction. On the other hand,
this freedom to locate the solution arbitrarily in $\phi$ is lost
in the Type-III solutions leading to a single pair of eigenvalues
at the origin of the spectral plane. In the case of the stable Type
II-out solutions reported above, e.g., for $\alpha=0.15$, there are
two additional eigenvalue pairs on the imaginary axis indicating
the spectral stability of the state (since real eigenvalues are
absent in this case). 
Aside from this branch of the form of II-out, all other identified
solutions have been found to be unstable.
A natural subsequent question then is that of the dynamical evolution of
these unstable solutions. 
After time propagation according to Eq.~(\ref{nls}) the unstable solutions mostly experience blowup in the case examples that we have
examined.  Such blowup can occur either on the inside or on the
outside of the torus.
This is determined predominantly by the nature of the solution,
but also from the nature of the associated perturbation.

\begin{figure}
\includegraphics[width=1.025\columnwidth]{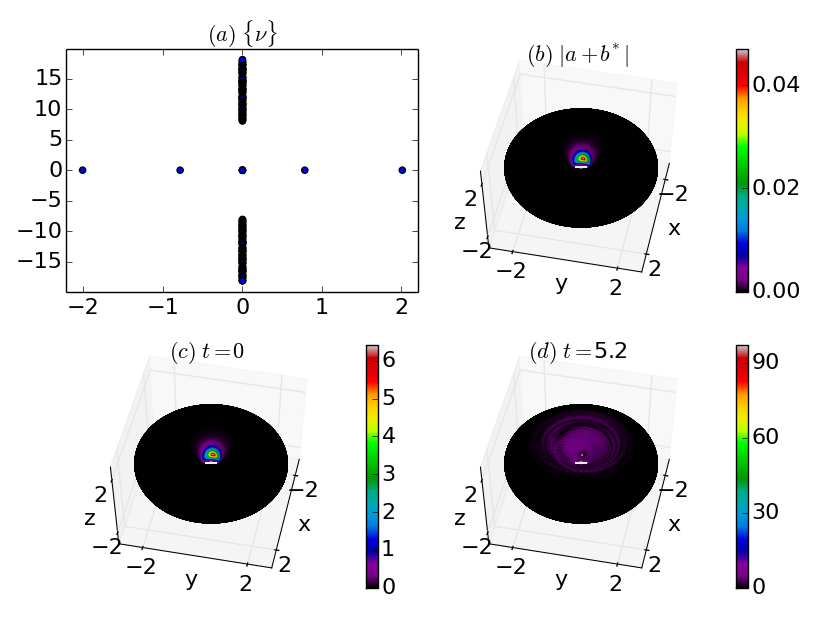} 
\caption{For $\alpha = 0.75$, $R = 1.6$ and $\mu = -8$ the time propagation of a stationary solution is demonstrated.  In (a) the eigenvalues $\{\nu\}$ are shown in the complex plane.  In (b) the magnitude of $a+b^*$ is shown on the torus according to the colorbar values, where $[a, b]^T$ is the eigenvector corresponding to the eigenvalue with max real part.  In (c)-(d) the stationary solution is shown at $t=0$ and at a later time $t=5.2$ when the solution approaches blowup. }
\label{rk_75mu-8}
\end{figure}

\begin{figure}
\includegraphics[width=1.025\columnwidth]{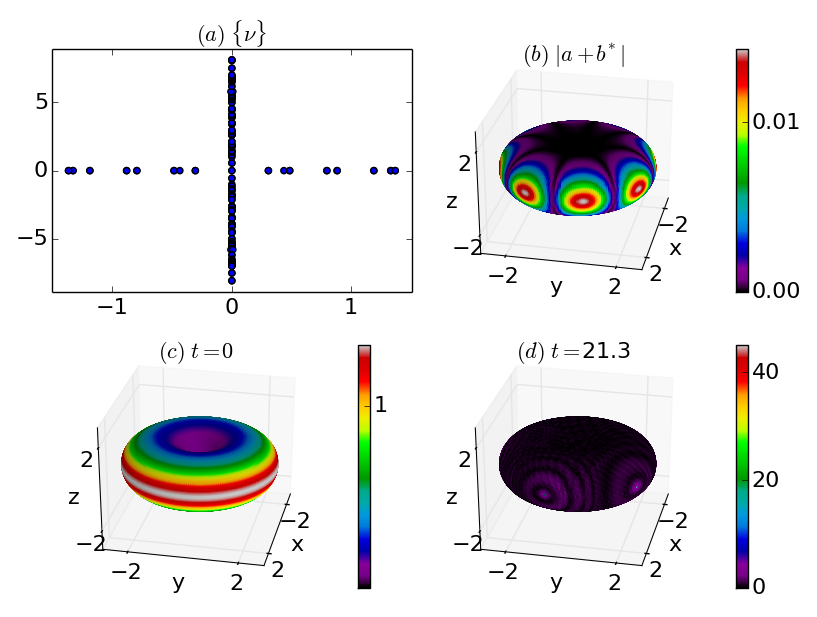} 
\caption{As in Figure \ref{rk_75mu-8} but for the values $\alpha = 0.75$, $R = 1.6$ and $\mu = -1$.  The stationary solution is shown at $t=0$ and at a later time $t=21.3$ when the solution approaches blowup at multiple spots around the exterior of the torus. }
\label{rk_75mu-1}
\end{figure}

\begin{figure}
\includegraphics[width=1.025\columnwidth]{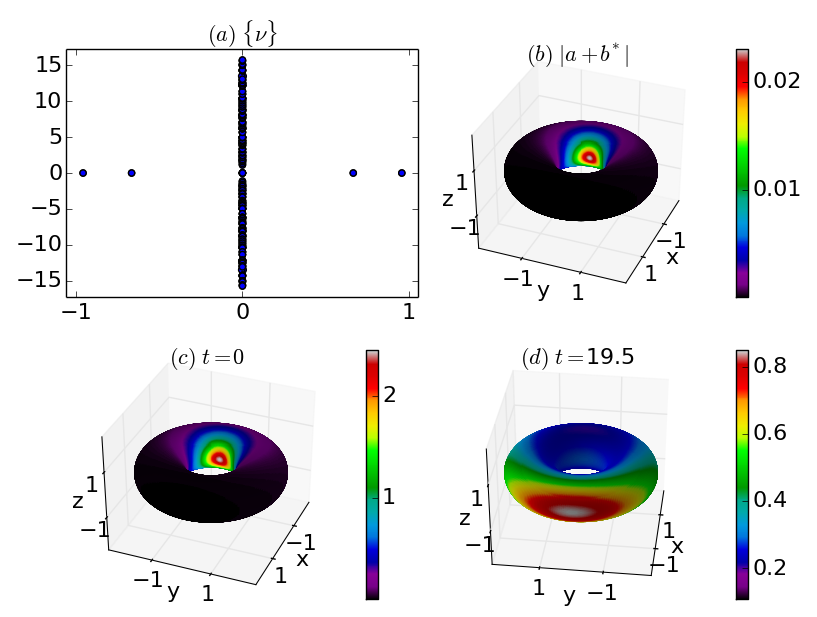} 
\caption{As in Figure \ref{rk_75mu-8} but for the values $\alpha = 0.5$, $R = 1.6$ and $\mu = -1$.  The stationary solution is shown at $t=0$ and at a later time $t=19.4$ when the solution mass moves to the exterior of the torus. {In this example a Type I-in solution approaches a Type II-out solution
    rather than being led to collapse as a result of its instability. } }
\label{rk_5}
\end{figure}

Examples of blowup are shown in Figures \ref{rk_75mu-8}-\ref{rk_75mu-1}.  In some cases a Type I-in or Type II-in unstable solution may approach one of the Type II-out stable solutions described above.  An example is shown in Figure \ref{rk_5}.
Such a phenomenology is the only alternative to collapsing
behavior that we have observed in the torus dynamics case examples that
we have considered.

\begin{figure}
\includegraphics[width=1.025\columnwidth]{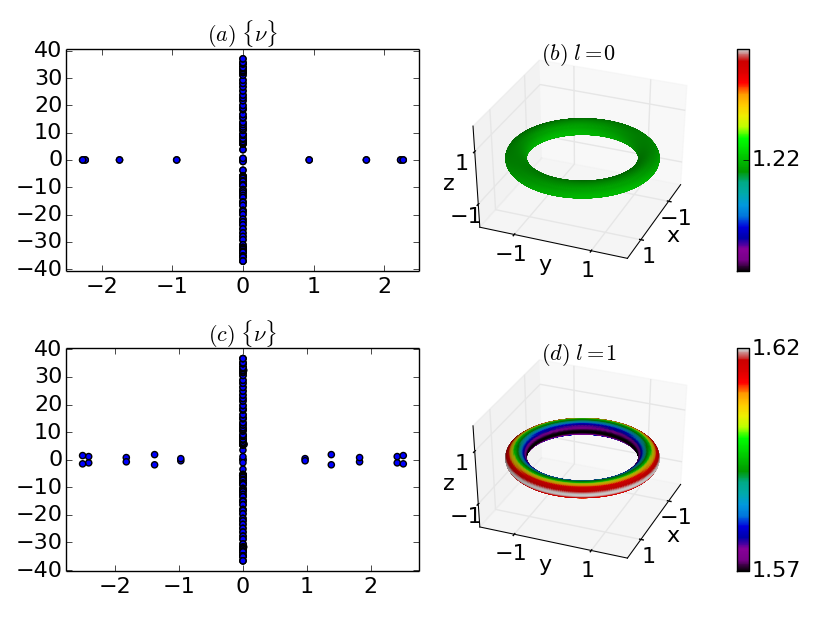} 
\caption{Plot (a) shows the eigenvalues of the solution for $l=0$ which is shown on the torus in (b).  Plot (c) similarly for $l=1$ in (d).  Both examples correspond to $\alpha = 0.15$ and $\mu = -2.36$.}
\label{expiphi}
\end{figure}

Lastly, we briefly consider the possibility of vortical solutions
bearing an additional phase factor of $e^{i l \phi}$. An example of
this type for the waveform of Type III is shown in Figure \ref{expiphi}.
From the left panels illustrating the spectral planes $(\nu_r,\nu_i)$
of eigenvalues $\nu=\nu_r+ i \nu_i$, we can infer that solutions with
$l=1$ are more unstable (via a larger number of unstable modes and
associated growth rates) than the $l=0$ ones. This has been typical
in the cases we have examined herein, hence we do not focus on these
vortical waveforms further.

\section{Conclusions and Outlook}

In the present work, we have set the stage for considering
nonlinear Schr{\"o}dinger equations and related dispersive
wave models on the torus. While in the mathematical community,
this is synonymous to a periodic boundary condition
system, as opposed to an infinite domain, here the details of
the geometry critically affect the linear (Laplacian) operator
adding a natural parameter to this system, namely the ratio of
minor over major axis of the torus. Furthermore, the explict
presence of the poloidal variable and the finite size periodicity
of the toroidal one induce drastic differences from the translationally
invariant two-dimensional case that is familiar in the NLS
realm~\cite{sulem,fibich}. Here, we encounter a situation where
(partially) translational as well as scale invariance are broken by the presence
of curvature leading to a potential spectral destabilization of
the resulting solutions. We have identified different types of states
including ones that are localized in both spatial directions,
localized in one and wrapping around (yet staying far from zero)
in the other, as well as localized
in one and homogeneous in the other. Among these, we found that
only the second type can have a range of frequencies (near the
linear limit) where the solutions are spectrally stable.
For too high frequencies and for all branches, nonlinearity takes
over and leads to collapse instabilities. However, sufficiently
close to the small amplitude limit, the interplay of curvature
and nonlinearity may dynamically lead to periodic
oscillations around a spectrally stable state, as shown in our
dynamics above.

Naturally, we feel that this first stab at this class of problems
opens numerous new directions to consider.
A very canonical one among them is to explore the possibility
of self-defocusing nonlinearities. On the one hand, these are
canonical for numerous atomic gases such as $^{87}Rb$ or
$^{23}Na$~\cite{stringari,pethick}
while on the other hand, they present the potential for fundamentally
distinct structures including ones bearing vorticity in the two-dimensional
realm. This vorticity could be in the form of localized vorticity
(point vortices), or in that of vorticity filaments, such as rings
wrapping potentially around either the toroidal or the poloidal direction.
Exploring such states and their stability has been a major theme
in atomic BECs~\cite{siambook} and its interplay with geometry
in this setting would present novel challenges and
potential outcomes. Yet another relevant possibility could be
to remain on the focusing interaction realm but ``negate''
the possibility of collapse by introducing photorefractive
nonlinearities as, e.g., in~\cite{njp_yang}.
In this case, the model at sufficiently high intensity returns
to its linear form and therefore collapse type instabilities
no longer occur. It would be especially relevant in this latter
setting to explore which among the different types of unstable solutions
identified herein become stabilized (or possibly vice versa).
These studies will be deferred to future work.

{\it Acknowledgements.} J.D.A. gratefully acknowledges computing support based on the Army Research Office ARO-DURIP Grant W911NF-15-1-0403.

\end{document}